\documentstyle[sprocl,epsf]{article}

 \catcode`@=11
\def\@cite#1#2{\unskip\nobreak\relax
    \def\@tempa{$\m@th^{\hbox{\the\scriptfont0 #1}}$}%
    \futurelet\@tempc\@citexx}
\def\@citexx{\ifx.\@tempc\let\@tempd=\@citepunct\else
    \ifx,\@tempc\let\@tempd=\@citepunct\else
    \let\@tempd=\@tempa\fi\fi\@tempd}
\def\@citepunct{\@tempc\edef\@sf{\spacefactor=\the\spacefactor\relax}\@tempa
    \@sf\@gobble}
 
\def\citenum#1{{\def\@cite##1##2{##1}\cite{#1}}}
\def\citea#1{\@cite{#1}{}}
 
\newcount\@tempcntc
\def\@citex[#1]#2{\if@filesw\immediate\write\@auxout{\string\citation{#2}}\fi
  \@tempcnta\z@\@tempcntb\m@ne\def\@citea{}\@cite{\@for\@citeb:=#2\do
    {\@ifundefined
       {b@\@citeb}{\@citeo\@tempcntb\m@ne\@citea\def\@citea{,}{\bf ?}\@warning
       {Citation `\@citeb' on page \thepage \space undefined}}%
    {\setbox\z@\hbox{\global\@tempcntc0\csname b@\@citeb\endcsname\relax}%
     \ifnum\@tempcntc=\z@ \@citeo\@tempcntb\m@ne
       \@citea\def\@citea{,}\hbox{\csname b@\@citeb\endcsname}%
     \else
      \advance\@tempcntb\@ne
      \ifnum\@tempcntb=\@tempcntc
      \else\advance\@tempcntb\m@ne\@citeo
      \@tempcnta\@tempcntc\@tempcntb\@tempcntc\fi\fi}}\@citeo}{#1}}
\def\@citeo{\ifnum\@tempcnta>\@tempcntb\else\@citea\def\@citea{,}%
  \ifnum\@tempcnta=\@tempcntb\the\@tempcnta\else
   {\advance\@tempcnta\@ne\ifnum\@tempcnta=\@tempcntb \else \def\@citea{--}\fi
    \advance\@tempcnta\m@ne\the\@tempcnta\@citea\the\@tempcntb}\fi\fi}

\catcode`@=12

\begin{document}

\title{\uppercase{Status of the AMANDA South Pole\\ Neutrino Detector}}

\author{\uppercase{Francis Halzen}\\
for the AMANDA Collaboration$^*$}

\address{}

\maketitle

\abstracts{
Initial deployment of optical modules near 1 and 2~kilometer depth indicate that deep polar ice is the most transparent known natural solid. Experience with early data has revealed that a detector, conceived to measure muons tracks, can also measure energy of high energy neutrinos as well as bursts of MeV neutrinos, e.g.\ produced by supernovae and gamma ray bursts. We plan to complete AMANDA this austral summer to form a detector of 11 deep strings instrumented over 400 meters height with 300 optical modules. We will argue that ice is the ideal medium to deploy a future kilometer-scale detector and discuss the first deployment of 10 strings of kilometer length.     
}

\section{Scientific Motivation}

The observed photon spectrum of conventional astronomy extends to energies in excess of 10~TeV, perhaps as high as 50~TeV. Astronomy can be done at much higher energy by measuring the arrival directions of cosmic rays with energy in excess of 100~EeV where their gyroradii exceed the size of our galaxy. More than 6 orders of magnitude in photon energy, or wavelength, are left unexplored. Neutrino telescopes have been conceived to fill this gap. Given the history of astronomy, it is difficult to imagine that this will be done without making major, and most likely totally surprising, discoveries. The 18 orders of magnitude in wavelength, from radio-waves to GeV gamma rays, are indeed sprinkled with unexpected discoveries. Neutrinos have the further advantage that, unlike photons of TeV energy and beyond, they are not absorbed by interstellar light.

Speculations that the highest energy photons and protons are produced by cosmic accelerators powered by the supermassive black holes at the center of active galaxies can be used to estimate the required effective volume of a neutrino telescope. The answer is 1~km$^3$\cite{halzen}. This estimate finds further support when studying the other diverse scientific missions of such an instrument which touch astronomy, astrophysics, cosmology, cosmic ray and particle physics. Model building suggests that the detection of  these accelerators may be within reach of much smaller detectors with effective area of order $10^4\rm\,m^2$\cite{halzen}. The AMANDA collaboration is ready to complete such an instrument within the next few months.

\break
At this meeting, the capability of neutrino telescopes to discover the particles that constitute the cold component of the dark matter is of special interest. The existence of the weakly interacting massive particles (WIMPs) can be inferred from observation of their annihilation products. Cold dark matter particles annihilate into neutrinos; {\it massive} ones will annihilate into {\it high-energy} neutrinos which can be detected in high-energy neutrino telescopes. This so-called indirect detection is greatly facilitated by the fact that the earth and the sun represent dense, nearby sources of accumulated cold dark matter particles. Galactic WIMPs, scattering off nuclei in the sun, lose energy. They may fall below escape velocity and be gravitationally trapped. Trapped WIMPs eventually come to equilibrium temperature and accumulate near the center of the sun. While the WIMP density builds up, their annihilation rate into lighter particles increases until equilibrium is achieved where the annihilation rate equals half of the capture rate. The sun has thus become a reservoir of WIMPs which we expect to annihilate mostly into heavy quarks and, for the heavier WIMPs, into weak bosons. The leptonic decays of the heavy quark and weak boson annihilation products turn the sun into a source of high-energy neutrinos with energies in the GeV to TeV range. Figure~1 displays the neutrino flux from the center of the earth calculated in the context of supersymmetric dark matter theories\cite{scopel}. The direct capture rate of the WIMPs in germanium detectors is shown for comparison. Contours indicate the parameter space favored by grand unified theories. Most of this parameter space can be covered by improving the capabilities of existing detectors by 2 orders of magnitude. For the indirect detection this is within reach of the AMANDA detector, given that at present a flux of 1~event per $100 \rm~m^2$ per year is excluded by the Baksan experiment\cite{suvorova}.

\begin{figure}[t]
\centering
\hspace{0in}\epsfxsize=4.5in\epsffile{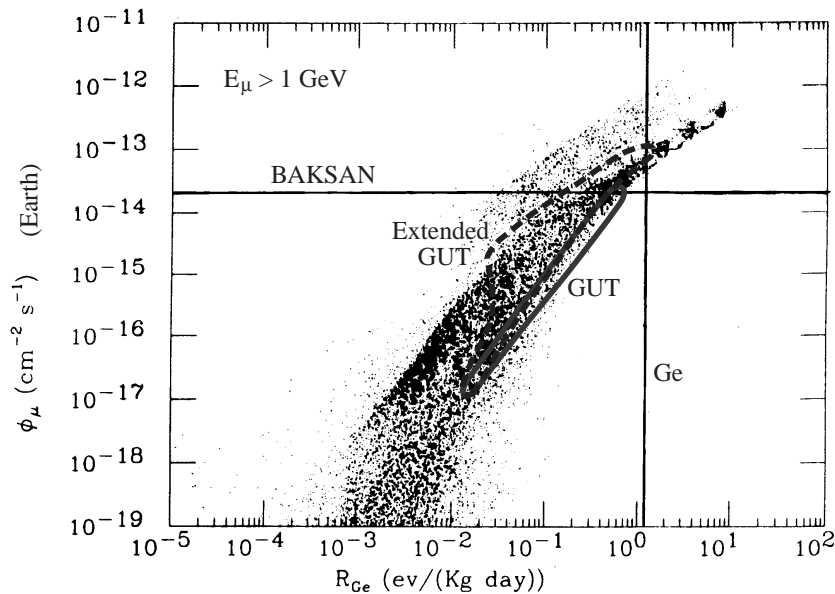}

\caption{Direct and indirect detection rates (for neutrinos from the center of the earth in the figure shown) of cold dark matter particles predicted by supersymmetric theory. Grand unified theories favor the parameter space indicated. Part of it is already excluded by present experiments as indicated by the horizontal and vertical lines.}
\end{figure}

\section{ AMANDA: Deployment and Technology}

The AMANDA project's goal is to commission a neutrino telescope by using natural Antarctic ice as a particle detector. With hot water, holes are melted into the 3~kilometer thick ice sheet at the South Pole. Optical modules (OMs), consisting of a photomultiplier in a pressure vessel and nothing else, are deployed and frozen into the ice. Hot water drilling progresses 1~centimeter every second; 2-kilometer deep holes are drilled in 4~days. The possibility to operate the detector with a data acquisition system placed right on top of the detector allows for a simple and non-hierarchical system where each OM is connected to the surface by its own cable. This provides high reliability against single point failure and makes evolutionary upgrades of the instrument possible. Furthermore, ice is a totally sterile medium devoid of radioactivity. Background counting rates in the OMs are extremely low: only a few hundred Hz for the deep AMANDA detector. This is to be compared with the 60~kHz background counting rate from natural radioactivity measured in the OMs used in the deep ocean experiments. The experiment can therefore be triggered by off-the-shelf electronics. It was indeed the low cost which represented the initial motivation for the experiment. The proponents argued that ice represented a competitive alternative to water even though the absorption length of blue light, relevant to the propagation of Cherenkov photons, was thought to be only 8~meters based on laboratory experiments. Nature has been more kind.

\begin{figure}%[t]
\centering
\hspace{0in}\epsfysize=6in\epsffile{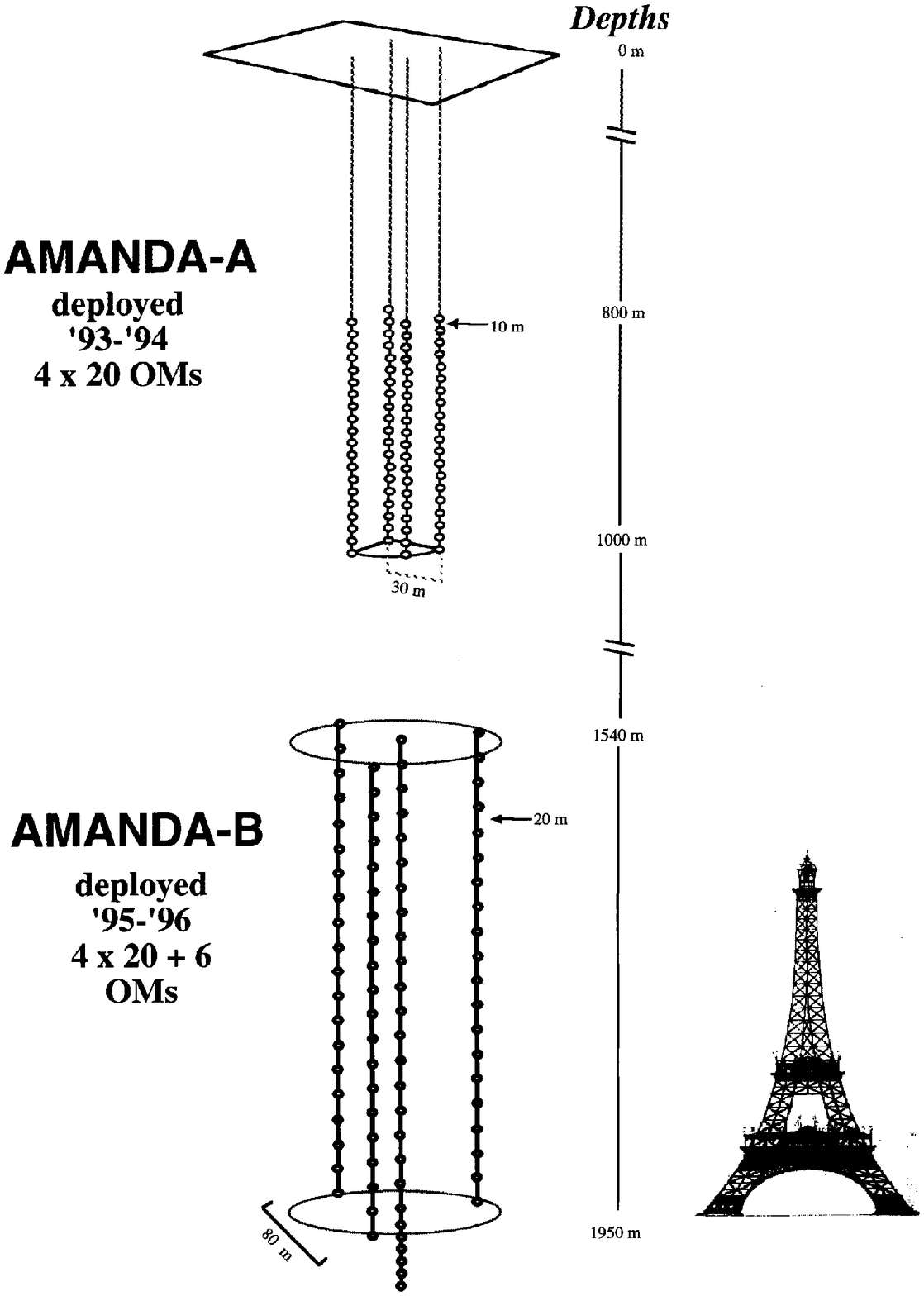}

\caption{}
\end{figure}

In Austral summer 93-94 the AMANDA A detector of 80 OMs was deployed on 4 strings positioned between 810~m and 1000~m; see Fig.~2. Not a single OM of the 73 which survived refreezing has failed since. The absorption length was found to be unexpectedly large, exceeding 200~m for wavelengths below 425~nm, more than one order of magnitude larger than laboratory ice\cite{science}. Remnant bubbles, much larger in size than expected, unfortunately limit the scattering length to tens of centimeters, preventing the reconstruction of muon tracks. With this set-back nature did however provide us with the hint that scattering can be exploited to measure energy. More about that later.

A deep, redesigned array, AMANDA B, has been partially deployed in bubble-free ice below 1500~m. The first 4 strings of 20 OMs, separated by 20~m along the string, were deployed in the 95-96 season. The signals from 14 stage Hamamatsu R5912-2 photomultipliers, operated at $10^9$ gain, are transmitted over coaxial cable. An arrival time resolution of 3~ns is achieved. As part of the evolving detector technology the signals of  an additional 6 OMs are transmitted over twisted pair cable which has yielded a factor 3 improvement in both rise-time and amplitude.  Using this technology, AMANDA B will be completed with 7 strings of 36 OMs in the 96-97 season. The smaller diameter of the cable makes possible the increased number of OMs per string. A new technology will be tested where the dynode signal of the photomultiplier drives a laser or a fast optical LED which transmits the analog signal over fiber optic cable. The same cable can be used which supplies laser calibration signals to a diffuser ball deployed along with every OM. It is likely that this method, if proven reliable, will be the AMANDA technology of the future. A pair of digital modules, produced by LBL and JPL, will also be tested as part of the next deployments.

The next stage of the experiment, AMANDA II, has been funded and will witness the first deployment of 10 strings of 1 kilometer length surrounding AMANDA B. It will be a zero-generation kilometer-scale detector from a technological point of view and, with an effective area likely to exceed $10^5$~m$^2$, sensitive to optimistic predictions for extraterrestrial neutrino fluxes.

\section{A Guided Tour of the AMANDA Data}

This is pioneering science --- surprises are routine. The two most important discoveries after operating AMANDA A and B for 2 years and 6 month, respectively, are that neutrino energy can be measured and that bursts of MeV-neutrinos, although well below the nominal threshold of the neutrino telescope, can be detected. We review the status of the data analysis, starting with the more traditional methods of operating the detector.     

\raggedbottom

\subsection{ Mapping of the Cherenkov cone from muons initiated by muon-neutrinos interacting inside or near the detector}

This is the traditional way one expected to operate a neutrino telescope. At TeV energy and above the direction of the muon and incident neutrino are aligned to better than 1$^\circ$, making astronomy possible. In order to implement this technique one must understand how light propagates in the Cherenkov medium. This is determined by measuring the propagation of Cherenkov photons radiated by cosmic ray muons and by independent methods. Notice however that one only has to determine {\em how} light propagates and does not have to understand {\em why} it propagates in the observed manner, the details of which may be very complicated.

AMANDA B registers atmospheric muons at a rate of 25~Hz by requiring a simple majority 8-fold majority trigger. This data is, on average, in agreement with Monte Carlo simulations which describe the propagation of Cherenkov photons in terms of time distributions obtained by studying the propagation of laser light pulsed into diffuser balls deployed with each OM. A YAG laser in the surface laboratory provides tunable pulses at 410 to 610~nm down the optical fiber to the diffuser. Most fibers exhibit a loss less than 10~dB over what is measured in the laboratory. Also a pulsed nitrogen laser (337~nm) at a depth of 1820~m, held at a temperature of 24$^\circ$~C, is operating flawlessly. Pulsed blue LED beacons with filters for 450 and 380~nm emission are operating at various depths. DC lamps at 350~nm, 380~nm, and broadband are also operating.

At 850~m depth, in ice formed just after the most recent ice age, the absorption length reaches a peak value of more than 300~m near the Cherenkov wavelength. Remnant bubbles limit the scattering length to less than 1~meter at depths less than 1~kilometer. Deeper AMANDA B measurements confirm the large absorption lengths observed at 1~kilometer and indicate a scattering length which is 2 orders of magnitude larger than for AMANDA A. The optical properties of the ice do not vary over the 500~m of depth below 1500~m over which the AMANDA B strings are deployed.  Using fast line fitting, rather than Cherenkov cone fitting of the muon tracks, a filter has been designed which can reduce, in real time, the 25~Hz rate to several hundred events a day which contain a significant fraction of the up-going neutrino events. These events can be transmitted via satellite on a daily basis. The efficiency of this filter has been demonstrated for 4 strings and will be installed once the detector is completed. The actual reconstruction of the Cherenkov signals takes into account the time delay of late photons due to scattering by maximum-likelihood fit. Monte Carlo simulations indicate that a resolution of 2.5~degrees can be achieved over an effective area of more than $10^4\rm\,m^2$ with the completed AMANDA B detector. The simulations demonstrate adequate rejection of down-going cosmic ray muons and led to a final design where one new string will be positioned at the center of the present 4-string array and 6 more strings with 36~OMs will be positioned on a radius of 60~m.  

\break
This raises the important point that reconstruction and background rejection of down-going cosmic ray muons is easier the larger the detector, especially because the scattering can now be exploited to measure muon energy-loss. A kilometer-scale ice detector is expected to operate as a directional Cherenkov detector as well as a total absorption calorimeter. The energy measurement is critical because, once it can be established that neutrino energy exceeds 10~TeV, cosmic origin of the signal is established because the atmospheric background rate is negligible as a result of the steeply falling spectrum; see Section~3.2. The problem of reconstructing muons in a kilometer-scale detector has been assessed experimentally by studying muon tracks registered in both the 1 and 2 kilometer-deep detectors. Such events are triggered at a rate of 0.1~Hz. Every 10 seconds a muon is tracked over 1.2 kilometer; a typical event is shown in Fig.~3. Below 1500~m the vertical muon triggers 2 strings separated by 79.5~m. The distance along the Cherenkov cone is over 100~m, yet, despite some evidence of scattering, the speed of light propagation of the track can be readily identified. We have already analysed $5 \times 10^5$  A-B coincidences with no evidence for any source of misreconstructed background events. Reconstruction of such tracks to a degree in bubble-free ice should not represent a challenge. 
   
\renewcommand{\thefigure}{\arabic{figure}a}
\begin{figure}%[t]
\centering
\hspace{0in}\epsfysize=6in\epsffile{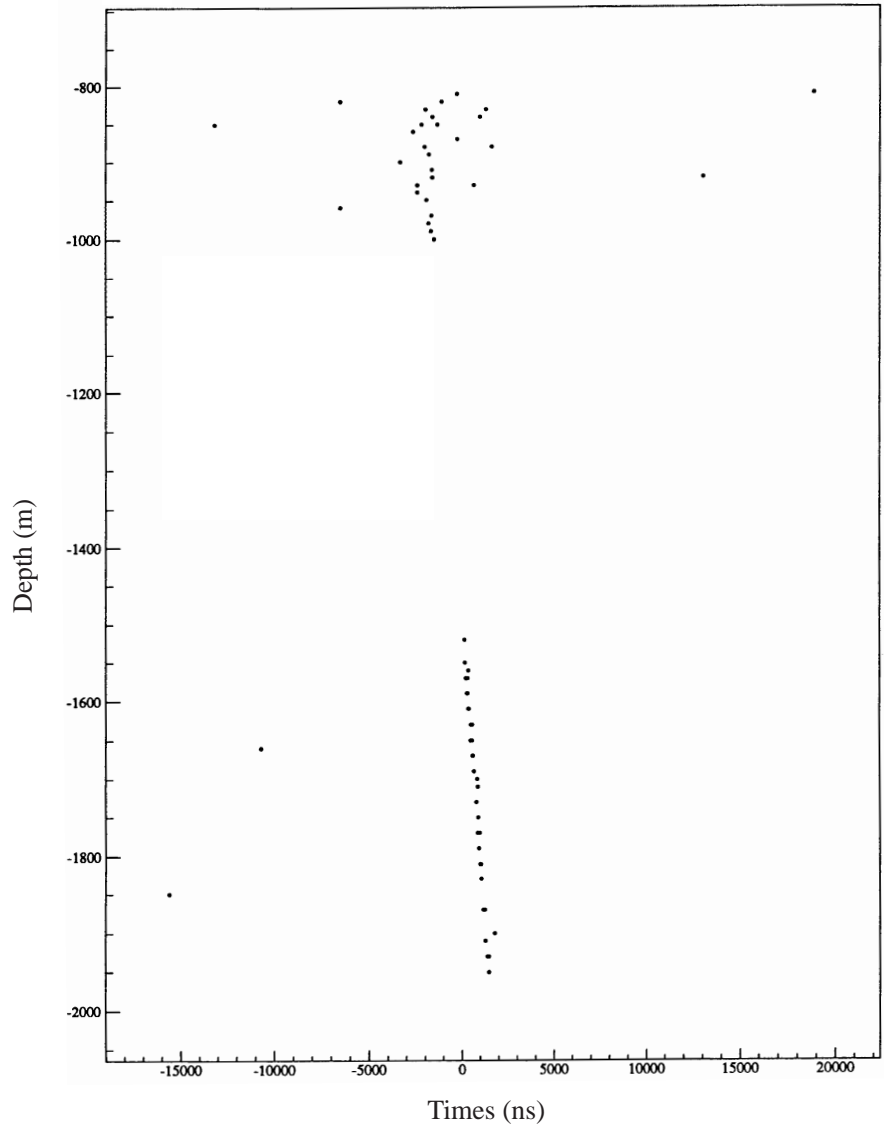}

\caption{Cosmic ray muon track triggered by both AMANDA A and B. Trigger times of the optical modules are shown as a function of depth. The diagram shows the diffusion of the track by bubbles above 1~km depth. Early and late hits, not associated with the track, are photomultiplier noise.}
\end{figure}

\addtocounter{figure}{-1}\renewcommand{\thefigure}{\arabic{figure}b}
\begin{figure}%[t]
\centering
\hspace{0in}\epsfysize=6in\epsffile{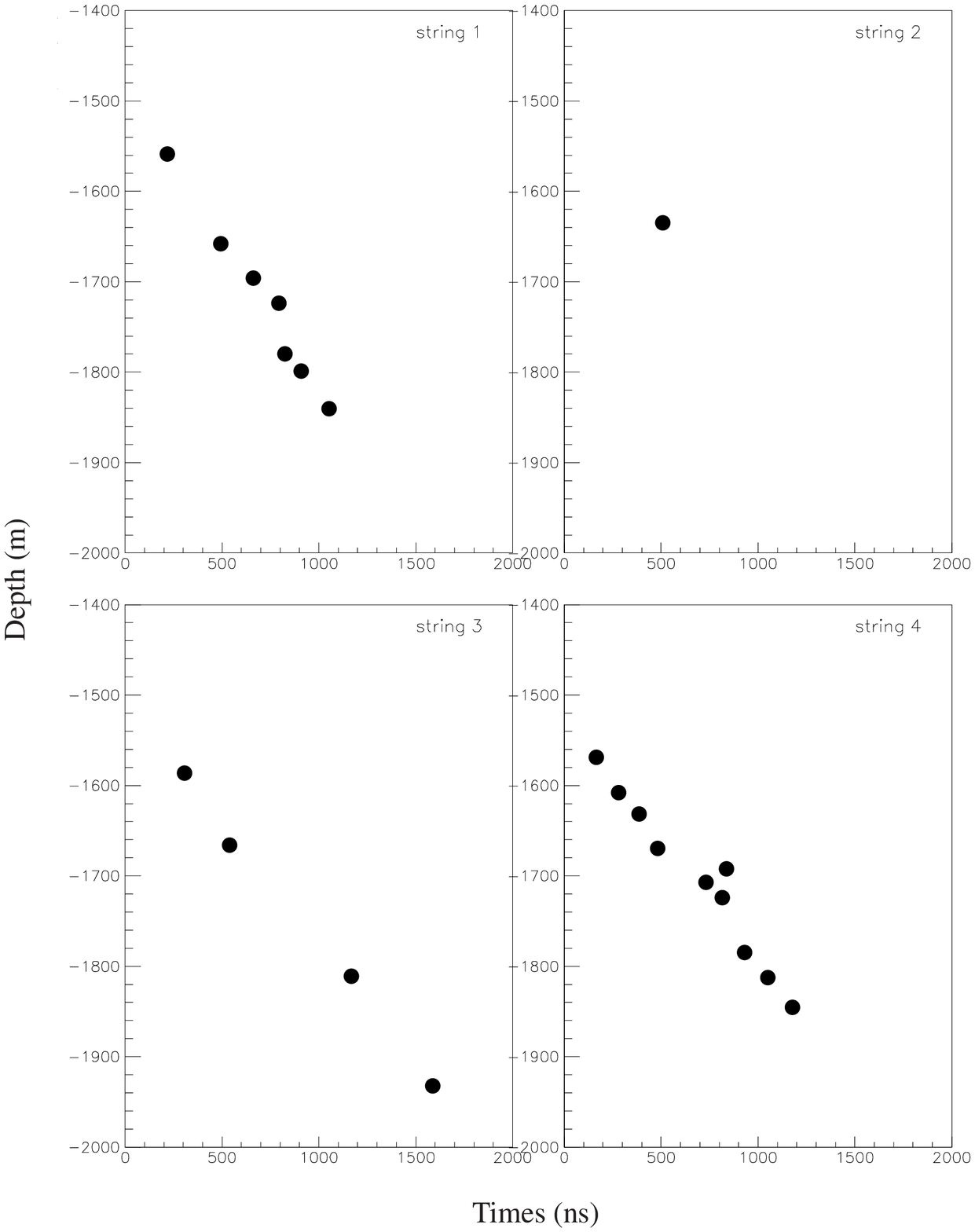}

\caption{Cosmic ray muon track triggered by both AMANDA A and B. Trigger times are shown separately for each string in the deep detector. In this event the muon mostly triggers OMs on strings 1 and 4 which are separated by 79.5~m. }
\end{figure}

\subsection{ Mapping of Electromagnetic Showers}

Ice is an ideal Cherenkov medium for a kilometer-scale neutrino detector. Simulations show that a PeV shower, initiated by a $\nu_e$ or $\nu_\tau$ or by catastrophic energy loss of a muon, travels over 500~meters in the very transparent ice. At this point the shower is isotropized by scattering and its energy can be determined. The first photon which arrives at an OM is not scattered by the ice provided it is within 200~meters from the point of origin of the shower. This will supply directional information. As already pointed out, once one establishes that the energy of the event exceeds those of the highest energy cosmic ray muons registered by the detector, the up-down discrimination problem is automatically solved. 

The shallow part of the AMANDA detector has been operated for over 2 years as a shower detector and the first year of data analysed. The atmospheric neutrino spectrum has been measured to several TeV. Background muons are observed up to 1~PeV. With a scattering length of order 50~cm, AMANDA A is an adequate total absorption calorimeter despite its limited size. Bubbles diffuse and contain the shower light very effectively. The shower can be mapped and its energy measured provided the point of origin is within 50~m of the instrumented ice. In bubbly ice the directional information is essentially lost.

The technology is also being developed to deploy a kilometer-scale detector in water. In a water detector a PeV shower will yield 5~$\sim$~10 times less detected photons and 100 times more noise hits. Even after removing all 1~p.e.\ signals which are totally dominated by background, half of the ${\sim}10^2$ OM signals from a PeV muon are produced by potassium decay in natural water\cite{stenger}. Increasing the noise in Fig.~3 by a factor 100 may not prevent one from ``seeing" the track. This is not the issue. The challenge is to trigger the event and to make sure that one is not confused by the noise more than once in $10^5$ times.  We should here also point out that whenever an OM registers 2 photons, the probability that both are scattered is reduced: the scattering length in ice at the 2 photoelectron level is significantly increased.

\subsection{Detection of bursts of MeV neutrinos (supernovae, gamma-ray bursts)}

An independent DAQ system has been installed which scans the counting rates in the OMs in time intervals of 5~microseconds to 0.5~seconds. The very low background counting rate of OMs deployed in the extremely transparent sterile ice is increased by a statistically significant amount by additional signals produced by bursts of MeV(!) neutrinos, e.g.\ from a supernova or, possibly, a gamma ray burst. A stellar collapse at the center of our galaxy will be observed with good statistical significance and the time profile of the neutrino emission determined with good statistics. Relative timing of neutrinos and gamma rays from a cosmological gamma ray burst would determine neutrino mass with a precision covering the range implied by the solar anomaly.

\section{Conclusions}

The AMANDA collaboration has deployed the necessary diagnostic tools to establish that ice, instrumented with photomultiplier tubes, can be an adequate particle detector. This program is nearing completion. A fall-back technology using OM's linked to the surface by coaxial cable has been demonstrated. Within a few months 300 OMs will be operating, providing us with the first opportunity to do science.

In the next season the first strings of kilometer length will be deployed. One should be able to commission a kilometer-scale detector over a period of 5 years for a price tag of 20$\sim$30 million dollars (including logistics!) without significantly expanding the scope of existing South Pole logistics.
 
\section*{\bf Acknowledgments}

We thank John Lynch for a careful reading of the manuscript. The AMANDA experiment is funded by the Physics and Polar Program Divisions of NSF, DESY, the Wallenberg Foundation (Sweden) and by the Universities of California and Wisconsin.

\bigskip
\noindent \llap{$^*$}The AMANDA Collaboration:\\
P.~Askebjer$^1$, S.W.~Barwick$^2$, R.~Bay$^6$, L.~Bergstr\"om$^1$, A.~Bouchta$^1$, S.~Carius$^3$. E.~Dahlberg$^1$, K.~Engel$^4$, B.~Erlandsson$^1$, A.~Goobar$^1$, L.~Gray$^4$, A.~Hallgren$^3$, F.~Halzen$^4$, H.~Heukenkamp$^4$, P.O.~Hulth$^1$, S.~Hundertmark$^5$, J.~Jacobsen$^4$, S.~Johansson$^1$, V.~Kandhadai$^4$, A.~Karle$^5$, I.~Liubarsky$^4$, D.~Lowder$^6$,\break
 T.~Mikolajski$^5$, T.C.~Miller$^4$, P.~Mock$^2$, R.~Morse$^4$, D.~Nygren$^8$, R.~Porrata$^2$, P.B.~Price$^6$, A.~Richards$^6$, H.~Rubinstein$^3$, E.~Schneider$^2$, C.~Spiering$^5$,\break
 O.~Streicher$^5$, Q.~Sun$^1$, T.~Thon$^5$, S.~Tilav$^4$, C.~Walck$^1$, C.~Wiebusch$^5$,\break
 R.~Wischnewski$^5$, G.~Yodh$^2$

\medskip\noindent
$^1$Dept.\ of Physics, Stockholm University, Sweden\\
$^2$Dept.\ of Physics, University of California, Irvine\\
$^3$Dept.\ of Physics, University of Uppsala, Sweden\\
$^4$Dept.\ of Physics, University of Wisconsin, Madison\\
$^5$DESY -- Inst.\ for High Energy Physics, Zeuthen, Germany\\
$^6$Dept.\ of Physics, University of California, Berkeley\\
$^7$Bartol Research Inst., University of Delaware, Newark\\
$^8$Lawrence Berkeley Laboratory, California


\bigskip

\section*{References}
\begin{thebibliography}{3}

\bibitem{halzen}
F. Halzen, {\it The case for a kilometer-scale neutrino detector}, in Nuclear and Particle Astrophysics and Cosmology, Proceedings of Snowmass~94, R.~Kolb and R.~Peccei, eds; T.~K.~Gaisser, F.~Halzen and T.~Stanev, {\it Physics Reports} {\bf 258}, 173 (1995).

\bibitem{scopel}
S.~Scopel, these proceedings.

\bibitem{suvorova}
O.~Suvorova, these proceedings; see also L.~Bergst\"rom, J.~Edsjo and P.~Gondolo, Phys.\ Rev.~D (to be published).

\bibitem{science}
The AMANDA collaboration, {\it Science} {\bf 267}, 1147 (1995)

\bibitem{stenger}
V.~Stenger, {\it Aspen Workshop on High Energy Neutrino Astronomy}, Aspen (1996).

\end{thebibliography}
\end{document}